\documentclass[11pt,twoside]{article}
\usepackage{asp2010}

\resetcounters

\begin{document}

\title{Quasar Structure Emerges from the Three Forms of Radiation Pressure}
\author{Martin Elvis,$^1$}
\affil{$^1$ Harvard-Smithsonian Center for Astrophysics, \\ 60 Garden
  St. Cambridge MA 02138 USA } 

\begin{abstract}
All quasar spectra show the same atomic features in the optical, UV, near-IR and
soft X-rays over all of cosmic time, luminosity black hole mass and accretion
rate. This is a puzzle.  Here I show that it is possible that all of these
atomic features can be accounted for by gas from an accretion disk driven the
three forms of radiation pressure: electron scattering, line driving and dust
driving.  The locations where they successfully drive an escaping wind, and
those where they produce only a failed wind are both needed.
\end{abstract}

\section{The Puzzle of Quasar Spectra}
It is remarkable that, to a good approximation, the spectra of all quasars and
active galactic nuclei (called quasars hereinafter) are the same over 13~Gyr of
cosmic time, 6 orders of magnitude in luminosity, and 3 orders of magnitude in
both black hole mass and Eddington ratio. This constancy poses a
problem. Whatever structures produce these atomic features they must be formed
by some robust physics that does not depend strongly on any of these parameters.
Accretion disk winds hint at a solution, but whether this can be realized has
not been clear.

The features are these: All show (1) permitted broad (FWHM$\sim$ 5000
km/s) emission lines (BELs); (2) narrow (FWHM$\sim$ 1000 km/s) emission lines
(NELs) that often extend into resolved bi-cones; (3) an Fe-K X-ray emission line
at 6.4 keV. Half or more show: high ionization (4) UV narrow (FWHM $<$200 km/s)
absorption lines (NALs) and (5) X-ray narrow (FWHM $<$300km/s) absorption lines
(known as 'Warm Absorbers, or "WAs"'), while about 15\% show (6) broad
($\Delta$v$\sim$ 10,000 km/s) absorbers, known as Broad Absorption Lines, or
"BALs").

There are two exceptions: in both "type 2" quasars and "blazars" some
or all of these features are covered up, either by reddening of the central
region (type 2s), or by overwhelming them with a beamed continuum from a
jet (blazars). But in type 2s the missing BELs can be seen in polarized light,
while in blazars a normal quasar spectrum can be glimpsed when the jet is
temporarily in a low state. So we believe that, intrinsically, all quasars have
the same features in their inner regions.

Here I show that it is possible that these atomic features can be completely
accounted for by the three forms of radiation pressure: electron scattering,
line driving and dust driving, both in the locations where they are effective in
driving an escaping wind, and where they produce only a failed wind.

\section{Funnel Wind Model for Quasars}

In 2000 I proposed a phenomenological model for quasar structure that could
account for all the atomic features in quasars (\cite{2000ApJ...545...63E},
'E00'). The model determined the geometry and kinematics of the quasar.
The model posits gas thrown off the accretion disk vertically in a narrow band
of radii. As this wind is hit by the continuum radiation it bends radially
outward, forming a funnel-shaped flow. BALs are seen looking along the radial
flow, which reaches $\tau_{es}\sim$1 so giving the Fe-K line. NALs and WAs are
seen looking almost perpendicularly through the flow, at low angles to the
accretion disk; viewed from above the wind no absorption features are seen. The
relative numbers of NALs, BALs and no absorption quasars set the angle of the
flow to the disk to be $\sim$30$^{\circ}$ [ignoring any possible equatorial
obscuration (see Elvis 2012, in prep.)]. The BELs are formed from a cool phase
of the wind.

\subsection{Funnel Wind Predictions and Tests}

A good model makes specific, testable, predictions. Table~1 lists predictions of
the E00 model, and the papers which tested them. In each case the model
survived the test, though some are still debated. 

\begin{table}
\caption{Predictions and Tests of the Elvis (2000) Funnel Wind Model}
\begin{small}
\begin{tabular}{|r|l|}
\hline
Ref \# & Prediction  \\
\hline
1 & WA is in outflow \\
2 & WA has narrow absorption lines \\
3 & WAs \& NALs have same outflow velocities  \\ 
4 & WAs \& NALs have consistent ionization states \& column densities  \\
5 & WAs \& NALs occur in the same objects  \\  
6 & NAL bi-cones will be hollow and matter bounded outflows \\ 
7 & WAs \& NALs are common in high luminosity objects  \\
8 & WAs \& NALs are common in edge-on AGNs  \\
9 & WA has a few distinct phases  \\
10& WA phases are in pressure balance  \\ 
11& WA is radially thin  \\
12& NALs arise at accretion disk radii  \\
13& WAs arise at accretion disk radii  \\
14& BELs have a large scale height  \\
15& BELs are dominated by rotation  \\
16& BAL regions exist in all quasars \\
17& BAL region rotates  \\
18& BAL scattering region creates narrow Fe-K emission  \\ 
\hline
\end{tabular}

{\footnotesize References for each test: (1) \cite{2000ApJ...535L..17K}; 
(2)\cite{2001ApJ...557....2C};
(3) \cite{2005ApJ...631..741G}, 
(4) \cite{1994ApJ...434..493M}, \cite{2000ApJ...536..101H},
    \cite{2003ApJ...597..832K};
(5) \cite{2001ASPC..224...45K};
(6) \cite{2000ApJ...532L.101C}, \cite{2011ApJ...727...71F};
(7) \cite{2005A&A...432...15P}, \cite{2003ApJ...599..116V};
(8) \cite{1997ApJ...489L..25L};
(9) \cite{2010ApJ...711..888A};
(10) \cite{2005ApJ...620..165K};
(11) \cite{2007ApJ...659.1022K};
(12) \cite{2005ApJ...623...85G}, \cite{2005ApJ...620..665A};
(13) \cite{2007ApJ...659.1022K}, \cite{2010ApJ...709..611D};
(14) \cite{2000ApJ...536..284K};
(15) \cite{2005MNRAS.359..846S};
(16) \cite{1999MNRAS.303..227Y};
(17) \cite{2007Natur.450...74Y};
(18) \cite{2008MNRAS.388..611S}.
}
\end{small}
\end{table}

The first 3 predictions were quickly confirmed by {\em Chandra} and {\em
XMM-Newton} grating spectra.  The 4th required the realization that partial
covering applied to the UV lines, while the 5th was demonstrated by explicitly
testing if NALs predicted WAs and vice versa. The 6th was demonstrated by {\em
Hubble} STIS long slit spectroscopy of narrow line region bi-cones. Number 7 was
shown with larger UV and X-ray samples, while \#~8 had already been shown, but
was not widely known.

Predictions~9 and 10 have been controversial. It is now clear that every good
X-ray spectrum {\em can} be fitted with a 2-3 phase WA in pressure equilibrium,
but more complex solutions are allowed.  Higher spectral resolution is needed to
be definitive. 

Tests 11-13 are closely connected. Initial results showed small radii, but a
number of newer observations clearly put certain NAL systems at large,
kiloparsec, distances. This is the greatest challenge so far to the funnel wind
model. How general this conclusion is remains unknown. 

BELs (\#14-15) have been less tested. Polarimetry indicates rotation in some
bright objects. BALs (\#17-18) are also poorly tested. Again, polarimetery shows
that BAL velocity gas is present in non-BAL objects and that the BAL region
rotates.

These results gave us the confidence to search for a physical understanding
based on this structure, one that can explain the more peculiar features in E00:
the thin wind region, and a flow neither polar nor equatorial.

\section{A Non-hydrodynamic Radiation Driving model}

Radiation can transfer momentum to matter in only three ways\footnotemark, via:
(1) electron scattering, (2) atomic absorption, (3) molecular/surface physics
absorption.  Normally we talk of these as Compton scattering, line driving and
dust driving. Below the Eddington limit gravity exceeds the Compton scattering
force, preventing a Compton scattered wind escaping; but few quasars reach
Eddington (\cite{2006ApJ...648..128K},\cite{2010MNRAS.402.2637S}).  Line driving
is hundreds of times more effective in O-stars, driving escaping winds well
below Eddington. A dust grain effectively absorbs all the incident light, and so
a medium with sufficient charged dust can even more effectively accelerate a
wind. Which of these mechanisms work in quasars?
\footnotetext{Neglecting pair production.}

\subsection{The QWIND Code}
We built a {\em non-hydrodynamic} code, QWIND (\cite{2010A&A...516A..89R},
'RE10').  This is justified as the gas is always moving supersonically. NAL/WA
blueshifts are $\sim$1000~km/s and the gas, at T$\le$10$^6$K, has a thermal
velocity of $\le$100~km/s, so this assumption is reasonable. Avoiding hydro let
us quickly explore parameter space.

QWIND assumes that packets of gas are ejected at {\em all} radii from the disk,
a generalization of E00, with some velocity and density, and follows their
equation of motion as they are irradiated (as in \cite{1980AJ.....85..329I},
\cite{1999PASJ...51..725W}).
CLOUDY is used to determine the ionization of the gas packet, and performs the
radiative transfer to filter the quasar continuum through the inner gas
packets. We used the \cite{1975ApJ...195..157C} 'CAK') formalism to determine
$\alpha(CAK)$, the multiplier, or gain, of line driving over pure Compton
scattering.  Importantly, the UV continuum has a cos$\theta$ dependence on angle
above the disk, while the X-ray continuum is considered isotropic. The UV/X-ray
ratio is an input parameter. Only the supermassive black hole gravity was
included.

For typical conditions QWIND produced the flow lines in Figure~1. A number of
distinct zones form naturally.

\begin{figure}
\centering
\includegraphics[width=2.5in, angle=-90]{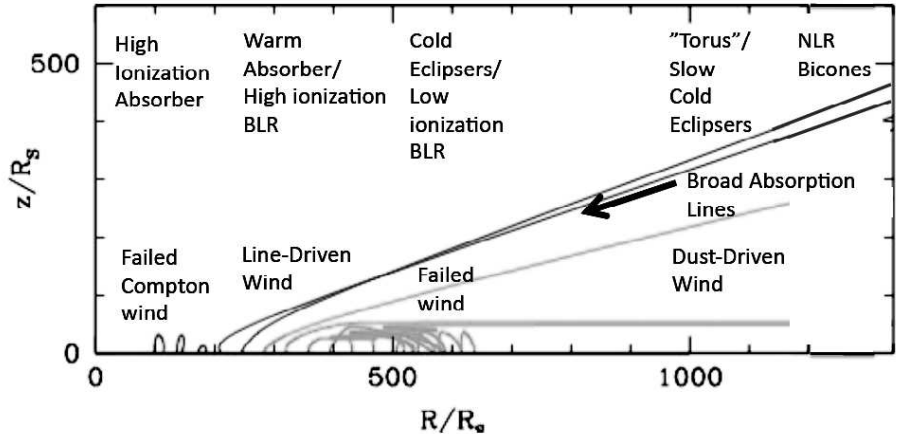}
\caption{\small Flowlines of gas from a quasar accretion disk generated by QWIND
  (annotated and expanded from \cite{2005ApJ...630L.129R}. Note the axes are
  linear in units of Schwartzchild radii.}
\end{figure}

\subsection{Compton Scattering Zone}
The innermost zone is overionized, so only Compton scattering is effective, but
below L(Edd) the wind cannot escape. Once $\tau_{es}>1$ this gas preferentially
scatters X-rays out of the disk plane, reducing the ionization parameter in the
next zone. This is the "hitchhiking gas" of \cite{1998ApJ...494..125M}). The gas
in this region still reaches a few 1000 km/s) before falling back. Such high
ionization, turbulent, variable velocity gas has been seen in NGC 1365
(\cite{2005ApJ...630L.129R}), with $\tau_{es}\sim$1.  Some L(Edd) objects show
high velocity winds (\cite{2003MNRAS.345..705P}, \cite{2002ApJ...579..169C}).

\subsection{Line Driving Zone}
The next zone out is where line driving becomes effective and an escaping wind
is formed.  Line driving has become widely accepted for NALs/WAs, following
Murray et al. (1995, \cite{1995ApJ...451..498M}). We find that in this zone
$\alpha(CAK)$ is only $\sim$10. The escaping wind still reaches BAL-like speeds
exceeding 10,000~km/s, but the low $\alpha(CAK)$ allows the wind to make a large
angle to the disk, contrary to Murray \& Chiang (1995
\cite{1995ApJ...454L.105M}), and consistent with E00.  Once the column
density reaches N$_H\sim$10$^{22}$cm$^{-2}$, the line driving UV photons are
absorbed, so this zone is physically thin, as in E00.  This N$_H$ is
characteristic of WAs.  High ionization BELs (CIV, OVI, NV, HeII, 'HiBELs') are
often considered part of a line-driven wind, and can produce the observed line
profiles (\cite{1997ApJ...474...91M}). Some multi-phase WA solutions include
a $\sim$10$^4$K branch, which may produce the HiBELs.  This should be explored
in detail.

\subsection{Failed Wind Zone}
Beyond the line driving zone the gas packets no longer reach escape velocity,
but do achieve a significant scale height.  Gas in this region could be
responsible for the low ionization BELs (FeII, MgII, 'LoBELs'), a modification
of E00. The LoBELs have long been located in the accretion disk
(\cite{1988MNRAS.232..539C}). This needs investigating using the LOC
formalism (\cite{1995ApJ...455L.119B}).  X-ray N$_H$ changes in days are
increasingly common (\cite{2004ApJ...615L..25E},
\cite{2007MNRAS.377..607P}), and have radii and densities suggestive of
LoBEL clouds.

\section{Dust Driven Zone}
RE10 do not include dust driving, as dust should be absent within the dust
sublimation radius, $r_{sub}$. However, \cite{2011A&A...525L...8C} point out
that the pressure and temperature conditions at which AGB star winds produce
dust are also found on quasar disk surfaces, from a few 1000 Schwartzchild radii
outward. Interior to $r_{sub}$ these clouds will initially be accelerated, but
their dust will soon evaporate, and they will cease to accelerate, probably
forming a failed wind. There are changes in X-ray N$_H$ on timescales of years
(e.g. Marinucci et al. in prep.), which might be caused by dusty clouds at these
radii crossing the line of sight. Beyond $r_{sub}$ these clouds will form a
wind, which may be one form of the obscuring "torus" needed to explain type~1
and type~2 quasars (Elvis 2012, in prep.).

\section{Line-Driven Wind Parameter Space}
Varying the parameters in QWIND lets us map out the conditions under which an
escaping wind forms. Clouds with BEL densities and temperatuers make a winds
hard to drive at high black hole mass; while an X-ray quiet continuum produces
winds over a much wider range of conditions, as expected
(\cite{1995ApJ...454L.105M}).

Interestingly high L/L(Edd) does not, as often assumed, imply faster winds.  At
higher L/L(Edd) the wind begins further out, and the 1/r$^2$ weakening of the
continuum lowers the final speed of the wind.

\section{Is Matter Launched at All Disk Radii?}
RE10 explicitly separate the problem of {\em accelerating} the clouds, which is
treated by QWIND, from the initial {\em launching} of clouds, by assuming that
launching occurs at all radii. There is no necessary connection between the two
processes. In fact, it seems unlikely that radiation driving alone can launch
material at all disk radii.

There is evidence that the disk supplies gas at many radii. In NGC~5548, the
H$\beta$ BEL radius 'breathes' in and out by a factor $\sim$3 as the $<$3500\AA\
luminosity changes by a factor $>$6 (\cite{1999ApJ...510..659P}). The few year
timescale is much shorter than the viscous timescale, but comparable to the
dynamical time. Plausibly there is pre-existing gas at all these radii, and the
continuum luminosity picks out the radius where the H$\beta$ emissivity is a
maximum, as in the LOC model, but now applied to a disk geometry. BELs form a
layered sequence from high ionization at small radii out to low ionization at
factor $\sim$20 larger radii (\cite{1999ApJ...521L..95P}, again as though the
gas is pre-existing and emitting whatever BEL is optimal.

The X-ray 'eclipses' by (probable) BEL clouds seen in NGC~1365, have a
comet-like form: small, dense heads with larger, lower density and higher
ionization, tails (\cite{2010A&A...517A..47M}).  The head lifetimes are
$\sim$2 months, substantially shorter than the orbital timescale.  The clouds
must then be constantly replenished. The disk is the only plausible source.

Magnetic reconnection could launch gas at all radii. Solar Coronal mass
ejections reach $\sim$1000~km/s, more than adequate for quasars. The
magneto-rotational instability (MRI), likely responsible for the viscosity in
quasar disks, can produce reconnection events. However, there may be a 'dead
zone' where the gas is too neutral to support MRI, as in proto-planetary disks
(\cite{2011ApJ...740L...6M}). Would there be a corresponding gap in the gas
supply for accelaration? MHD winds in general are another means of accelerating
quasar winds (\cite{2008ApJ...685..160N}). Different acceleration processes
may act in different quasars.  Tests are needed to discriminate between
them.

\section{Conclusions}
There is an elegance to this view of quasar structure. To form so many observed
phenomena from such simple underlying physics is economical. That initially
strange features of the funnel wind model are natural consequences of the onset
and failure of the three forms of radiation driving is particularly appealing.

This physics is robust to changes in parameters, including L/L(Edd) and black
hole mass, as boundary conditions are not important.  Anything beyond the
immediate environs of the black hole does not change the wind physics, so cosmic
time is not important. Hence the puzzle of quasar spectra, with which I began,
can be understood: all quasars are much the same because radiative acceleration
in a disk setting is always the same.

It will now be important to build on this success by seeing if the model can
produce second-order effects, such as eigenvector 1
(\cite{1992ApJS...80..109B}), and whether a predictive model for the energy and
momentum carried by the wind can be produced, in order to have a physical model
of quasar feedback.

\acknowledgements The author would especially like to the thank Guido Risaliti,
and also Bozena Czerny for discussions of her work. He is grateful for the
opportunity to present this work in Charleston.

\bibliography{ELVIS}

\begin{thebibliography}{}
\expandafter\ifx\csname natexlab\endcsname\relax\def\natexlab#1{#1}\fi
\expandafter\ifx\csname url\endcsname\relax
  \def\url#1{\texttt{#1}}\fi
\expandafter\ifx\csname urlprefix\endcsname\relax\def\urlprefix{URL }\fi
\providecommand{\eprint}[2][]{\url{#2}}

\bibitem[{{Andrade-Vel{\'a}zquez} \& {et al.}(2010)}]{2010ApJ...711..888A}
{Andrade-Vel{\'a}zquez}, M., \& {et al.} 2010, \apj, 711, 888

\bibitem[{{Arav} et~al.(2005){Arav}, {Kaastra}, {Kriss}, {Korista}, {Gabel}, \&
  {Proga}}]{2005ApJ...620..665A}
{Arav}, N., {Kaastra}, J., {Kriss}, G.~A., {Korista}, K.~T., {Gabel}, J., \&
  {Proga}, D. 2005, \apj, 620, 665

\bibitem[{{Baldwin} et~al.(1995){Baldwin}, {Ferland}, {Korista}, \&
  {Verner}}]{1995ApJ...455L.119B}
{Baldwin}, J., {Ferland}, G., {Korista}, K., \& {Verner}, D. 1995, \apjl, 455,
  L119

\bibitem[{{Boroson} \& {Green}(1992)}]{1992ApJS...80..109B}
{Boroson}, T.~A., \& {Green}, R.~F. 1992, \apjs, 80, 109

\bibitem[{{Castor} et~al.(1975){Castor}, {Abbott}, \&
  {Klein}}]{1975ApJ...195..157C}
{Castor}, J.~I., {Abbott}, D.~C., \& {Klein}, R.~I. 1975, \apj, 195, 157

\bibitem[{{Chartas} et~al.(2002){Chartas}, {Brandt}, {Gallagher}, \&
  {Garmire}}]{2002ApJ...579..169C}
{Chartas}, G., {Brandt}, W.~N., {Gallagher}, S.~C., \& {Garmire}, G.~P. 2002,
  \apj, 579, 169

\bibitem[{{Collin-Souffrin} et~al.(1988){Collin-Souffrin}, {Dyson}, {McDowell},
  \& {Perry}}]{1988MNRAS.232..539C}
{Collin-Souffrin}, S., {Dyson}, J.~E., {McDowell}, J.~C., \& {Perry}, J.~J.
  1988, \mnras, 232, 539

\bibitem[{{Collinge} \& {et al.}(2001)}]{2001ApJ...557....2C}
{Collinge}, M.~J., \& {et al.} 2001, \apj, 557, 2

\bibitem[{{Crenshaw} \& {Kraemer}(2000)}]{2000ApJ...532L.101C}
{Crenshaw}, D.~M., \& {Kraemer}, S.~B. 2000, \apjl, 532, L101

\bibitem[{{Czerny} \& {Hryniewicz}(2011)}]{2011A&A...525L...8C}
{Czerny}, B., \& {Hryniewicz}, K. 2011, \aap, 525, L8

\bibitem[{{Dunn} \& {et al.}(2010)}]{2010ApJ...709..611D}
{Dunn}, J.~P., \& {et al.} 2010, \apj, 709, 611

\bibitem[{{Elvis}(2000)}]{2000ApJ...545...63E}
{Elvis}, M. 2000, \apj, 545, 63. \eprint{arXiv:astro-ph/0008064}

\bibitem[{{Elvis} et~al.(2004){Elvis}, {Risaliti}, {Nicastro}, {Miller},
  {Fiore}, \& {Puccetti}}]{2004ApJ...615L..25E}
{Elvis}, M., {Risaliti}, G., {Nicastro}, F., {Miller}, J.~M., {Fiore}, F., \&
  {Puccetti}, S. 2004, \apjl, 615, L25

\bibitem[{{Fischer} et~al.(2011){Fischer}, {Crenshaw}, {Kraemer}, {Schmitt},
  {Mushotsky}, \& {Dunn}}]{2011ApJ...727...71F}
{Fischer}, T.~C., {Crenshaw}, D.~M., {Kraemer}, S.~B., {Schmitt}, H.~R.,
  {Mushotsky}, R.~F., \& {Dunn}, J.~P. 2011, \apj, 727, 71

\bibitem[{{Gabel} \& {et al.}(2005{\natexlab{a}})}]{2005ApJ...631..741G}
{Gabel}, J.~R., \& {et al.} 2005{\natexlab{a}}, \apj, 631, 741

\bibitem[{{Gabel} \& {et al.}(2005{\natexlab{b}})}]{2005ApJ...623...85G}
--- 2005{\natexlab{b}}, \apj, 623, 85

\bibitem[{{Hamann} et~al.(2000){Hamann}, {Netzer}, \&
  {Shields}}]{2000ApJ...536..101H}
{Hamann}, F.~W., {Netzer}, H., \& {Shields}, J.~C. 2000, \apj, 536, 101

\bibitem[{{Icke}(1980)}]{1980AJ.....85..329I}
{Icke}, V. 1980, \aj, 85, 329

\bibitem[{{Kaspi} \& {et al}(2000)}]{2000ApJ...535L..17K}
{Kaspi}, S., \& {et al} 2000, \apjl, 535, L17

\bibitem[{{Kollmeier} \& {et al.}(2006)}]{2006ApJ...648..128K}
{Kollmeier}, J.~A., \& {et al.} 2006, \apj, 648, 128.
  \eprint{arXiv:astro-ph/0508657}

\bibitem[{{Korista} \& {Goad}(2000)}]{2000ApJ...536..284K}
{Korista}, K.~T., \& {Goad}, M.~R. 2000, \apj, 536, 284

\bibitem[{{Kriss}(2001)}]{2001ASPC..224...45K}
{Kriss}, G.~A. 2001, in Probing the Physics of Active Galactic Nuclei, edited
  by {B.~M.~Peterson, R.~W.~Pogge, \& R.~S.~Polidan}, vol. 224 of Astronomical
  Society of the Pacific Conference Series, 45

\bibitem[{{Krongold} \& {et al.}(2007)}]{2007ApJ...659.1022K}
{Krongold}, Y., \& {et al.} 2007, \apj, 659, 1022

\bibitem[{{Krongold} et~al.(2003){Krongold}, {Nicastro}, {Brickhouse}, {Elvis},
  {Liedahl}, \& {Mathur}}]{2003ApJ...597..832K}
{Krongold}, Y., {Nicastro}, F., {Brickhouse}, N.~S., {Elvis}, M., {Liedahl},
  D.~A., \& {Mathur}, S. 2003, \apj, 597, 832

\bibitem[{{Krongold} et~al.(2005){Krongold}, {Nicastro}, {Elvis}, {Brickhouse},
  {Mathur}, \& {Zezas}}]{2005ApJ...620..165K}
{Krongold}, Y., {Nicastro}, F., {Elvis}, M., {Brickhouse}, N.~S., {Mathur}, S.,
  \& {Zezas}, A. 2005, \apj, 620, 165

\bibitem[{{Leighly} et~al.(1997){Leighly}, {Mushotzky}, {Nandra}, \&
  {Forster}}]{1997ApJ...489L..25L}
{Leighly}, K.~M., {Mushotzky}, R.~F., {Nandra}, K., \& {Forster}, K. 1997,
  \apjl, 489, L25

\bibitem[{{Maiolino} \& {et al.}(2010)}]{2010A&A...517A..47M}
{Maiolino}, R., \& {et al.} 2010, \aap, 517, A47

\bibitem[{{Martin} \& {Lubow}(2011)}]{2011ApJ...740L...6M}
{Martin}, R.~G., \& {Lubow}, S.~H. 2011, \apjl, 740, L6

\bibitem[{{Mathur} et~al.(1994){Mathur}, {Wilkes}, {Elvis}, \&
  {Fiore}}]{1994ApJ...434..493M}
{Mathur}, S., {Wilkes}, B., {Elvis}, M., \& {Fiore}, F. 1994, \apj, 434, 493

\bibitem[{{Murray} \& {Chiang}(1995)}]{1995ApJ...454L.105M}
{Murray}, N., \& {Chiang}, J. 1995, \apjl, 454, L105

\bibitem[{{Murray} \& {Chiang}(1997)}]{1997ApJ...474...91M}
--- 1997, \apj, 474, 91

\bibitem[{{Murray} \& {Chiang}(1998)}]{1998ApJ...494..125M}
--- 1998, \apj, 494, 125

\bibitem[{{Murray} et~al.(1995){Murray}, {Chiang}, {Grossman}, \&
  {Voit}}]{1995ApJ...451..498M}
{Murray}, N., {Chiang}, J., {Grossman}, S.~A., \& {Voit}, G.~M. 1995, \apj,
  451, 498

\bibitem[{{Nenkova} et~al.(2008){Nenkova}, {Sirocky}, {Nikutta}, {Ivezi{\'c}},
  \& {Elitzur}}]{2008ApJ...685..160N}
{Nenkova}, M., {Sirocky}, M.~M., {Nikutta}, R., {Ivezi{\'c}}, {\v Z}., \&
  {Elitzur}, M. 2008, \apj, 685, 160

\bibitem[{{Peterson} \& {et al.}(1999)}]{1999ApJ...510..659P}
{Peterson}, B.~M., \& {et al.} 1999, \apj, 510, 659

\bibitem[{{Peterson} \& {Wandel}(1999)}]{1999ApJ...521L..95P}
{Peterson}, B.~M., \& {Wandel}, A. 1999, \apjl, 521, L95

\bibitem[{{Piconcelli} et~al.(2005){Piconcelli}, {Jimenez-Bail{\'o}n},
  {Guainazzi}, {Schartel}, {Rodr{\'{\i}}guez-Pascual}, \&
  {Santos-Lle{\'o}}}]{2005A&A...432...15P}
{Piconcelli}, E., {Jimenez-Bail{\'o}n}, E., {Guainazzi}, M., {Schartel}, N.,
  {Rodr{\'{\i}}guez-Pascual}, P.~M., \& {Santos-Lle{\'o}}, M. 2005, \aap, 432,
  15

\bibitem[{{Pounds} et~al.(2003){Pounds}, {Reeves}, {King}, {Page}, {O'Brien},
  \& {Turner}}]{2003MNRAS.345..705P}
{Pounds}, K.~A., {Reeves}, J.~N., {King}, A.~R., {Page}, K.~L., {O'Brien},
  P.~T., \& {Turner}, M.~J.~L. 2003, \mnras, 345, 705

\bibitem[{{Puccetti} et~al.(2007){Puccetti}, {Fiore}, {Risaliti}, {Capalbi},
  {Elvis}, \& {Nicastro}}]{2007MNRAS.377..607P}
{Puccetti}, S., {Fiore}, F., {Risaliti}, G., {Capalbi}, M., {Elvis}, M., \&
  {Nicastro}, F. 2007, \mnras, 377, 607

\bibitem[{{Risaliti} \& {Elvis}(2010)}]{2010A&A...516A..89R}
{Risaliti}, G., \& {Elvis}, M. 2010, \aap, 516, A89. \eprint{0911.0958}

\bibitem[{{Risaliti} \& {et al}(2005)}]{2005ApJ...630L.129R}
{Risaliti}, G., \& {et al} 2005, \apjl, 630, L129

\bibitem[{{Sim} et~al.(2008){Sim}, {Long}, {Miller}, \&
  {Turner}}]{2008MNRAS.388..611S}
{Sim}, S.~A., {Long}, K.~S., {Miller}, L., \& {Turner}, T.~J. 2008, \mnras,
  388, 611

\bibitem[{{Smith} et~al.(2005){Smith}, {Robinson}, {Young}, {Axon}, \&
  {Corbett}}]{2005MNRAS.359..846S}
{Smith}, J.~E., {Robinson}, A., {Young}, S., {Axon}, D.~J., \& {Corbett}, E.~A.
  2005, \mnras, 359, 846

\bibitem[{{Steinhardt} \& {Elvis}(2010)}]{2010MNRAS.402.2637S}
{Steinhardt}, C.~L., \& {Elvis}, M. 2010, \mnras, 402, 2637. \eprint{0911.1355}

\bibitem[{{Vestergaard}(2003)}]{2003ApJ...599..116V}
{Vestergaard}, M. 2003, \apj, 599, 116

\bibitem[{{Watarai} \& {Fukue}(1999)}]{1999PASJ...51..725W}
{Watarai}, K.-y., \& {Fukue}, J. 1999, \pasj, 51, 725

\bibitem[{{Young} et~al.(2007){Young}, {Axon}, {Robinson}, {Hough}, \&
  {Smith}}]{2007Natur.450...74Y}
{Young}, S., {Axon}, D.~J., {Robinson}, A., {Hough}, J.~H., \& {Smith}, J.~E.
  2007, \nat, 450, 74

\bibitem[{{Young} et~al.(1999){Young}, {Corbett}, {Giannuzzo}, {Hough},
  {Robinson}, {Bailey}, \& {Axon}}]{1999MNRAS.303..227Y}
{Young}, S., {Corbett}, E.~A., {Giannuzzo}, M.~E., {Hough}, J.~H., {Robinson},
  A., {Bailey}, J.~A., \& {Axon}, D.~J. 1999, \mnras, 303, 227

\end{thebibliography}

\end{document}